\documentclass[authoryear,preprint,review,12pt]{elsarticle}
\usepackage{natbib}
\usepackage{graphicx}
\usepackage{epsfig,subfigure}
\usepackage{amssymb}
\usepackage{amsthm}
\usepackage{amsmath}
\def\astrobj#1{#1}
\journal{New Astronomy}
\begin{document}
\begin{frontmatter}
\title{Rapid high-amplitude $\gamma$-ray variability in Blazar PKS 0507+17}
\author[1,2,3]{Liao N.H.\corref{cor1}}
\ead{liaonh@ynao.ac.cn}
\author[1,2]{Bai J.M. \corref{cor2}}

\address[1]{Yunnan Observatories, Chinese Academy of Sciences, Kunming 650011, China}
\address[2]{Key Laboratory for the Structure and Evolution of Celestial Objects, Chinese Academy of Sciences}
\address[3]{University of Chinese Academy of Sciences, Beijing 100049, China}

\begin{abstract}
We report a detailed analysis of $\gamma$-ray data of \astrobj{2FGL J0509.9+1802} observed by the Large Area Telescope on board {\it Fermi} satellite, especially focusing on April 2013 when extraordinary $\gamma$-ray variability has been detected. Localization of $\gamma$-ray emission during this epoch suggests that the $\gamma$-ray source only associates with \astrobj{PKS 0507+17}. The $\gamma$-ray emission of PKS 0507+17 is identified at the first time. The daily peak flux is over two orders of magnitude higher than the first two-year average flux, giving an isotropic $\gamma$-ray luminosity of $\simeq4\times10^{48}$ erg $\rm s^{-1}$. Rapid $\gamma$-ray variability with doubling time of 2-3 hours has been detected. Such a short doubling timescale has been detected for only a few bright blazars and indicates a location of $\gamma$-ray emission inside the broad line region. Together with the bluer-when-brighter $\gamma$-ray spectra, the variability phenomena could be well explained by the classic flat-spectrum radio quasar variability model that includes a fast injection of accelerated electrons and the external Compton cooling process.
\end{abstract}


\end{frontmatter}


\section{Introduction}

Blazars, including flat-spectrum radio quasars (FSRQs) and BL Lacertae objects (BL Lac objects), is a subclass of  radio-loud active galactic nuclei (AGNs; Urry \& Padovani 1995). Their luminous, rapidly variable, and polarized non-thermal continuum emissions are widely accepted to be produced in the relativistic jets oriented close to the line of sight (Blandford \& Rees 1978; Ulrich et al. 1997). Their broadband electromagnetic radiations extend from radio to $\gamma$-ray energies. Blazars emit their most energies in the $\gamma$-ray domain, dominating the extragalactic $\gamma$-ray sky (Nolan et al. 2012). However, due to the unresolved emission regions, the location of the $\gamma$-ray emission is poorly known, which constitutes an obstacle to understand the energy dissipation in blazars.

Significant $\gamma$-ray variability is a distinct character for blazars. Extremely rapid $\gamma$-ray variability, such that flux doubling time is a few minutes, has been detected for four blazars in the TeV domain by the ground-based Cerenkov telescopes (Aharonian et al. 2007; Albert et al. 2007; Aleksi{\'c} et al. 2011; Arlen et al. 2013). These episodes put severe challenge on the classic one-zone homogeneous leptonic blazar radiation models that include synchrotron and inverse Compton (IC) process (Begelman et al. 2008). In the GeV range, there are only two FSRQs, \astrobj{PKS 1622-29} (Mattox et al. 1997) and \astrobj{3C 279} (Wehrle et al. 1998), from which flux variations on several hours have been detected by CGRO/EGRET. Recently, thanks to the operation of the Large Area Telescope (LAT) on {\it Fermi} $\gamma$-Ray Space Telescope (Atwood et al. 2009), number of blazars with detection of flux variability on the sub-daily timescales is increasing (e.g. Foschini et al. 2011; Vovk \& Neronov 2013). Fast $\gamma$-ray variability together with the absorption features in $\gamma$-ray spectra (Poutanen \& Stern 2010), suggest that the dissipation region is probably within the Broad Line Region (BLR). However, detailed multi-wavelength studies including correlation and lag analyses on outbursts, together with kinematic studies on Very Long Baseline Interferometry (VLBI) component ejection, suggest that the dissipation site of jet locates at pc scale (e.g. Marscher et al. 2008; Agudo et al. 2011).

The second LAT AGN catalog (2LAC) lists 1017 high Galactic latitudes $\gamma$-ray sources associated with AGNs (Ackermann et al. 2011). The clean sample of 2LAC includes 886 AGNs which are classified by single counterparts and without analysis flags. Due to the relatively poor angular resolution of LAT, (68\% containment angle) better than $1^{\circ}$ at energies above 1~GeV, there are $\gamma$-ray sources corresponding to multiple counterparts. For \astrobj{2FGL J0509.9+1802}, there are two radio counterparts with separation angle of only $\sim0.1^{\circ}$ both falling into the 95\% confidential level (C.L.) containment angle (Ackermann et al. 2011). These two radio counterparts demonstrate long jets in Very Long Baseline Array images (Linford et al. 2012). One of the radio counterparts, \astrobj{CRATES J0509+1806}, is listed in the CRATES catalog as a flat spectrum radio source (Healey et al. 2007). The other radio counterpart is known as \astrobj{PKS 0507+17}.

\astrobj{PKS 0507+17} is a bright well studied radio source. It was identified as an extragalactic compact radio source by VLBI observation (Wehrle et al. 1984). The source was collected in the Deep X-Ray Radio Blazar Survey and indentified as a FSRQ with redshift of 0.416 (Perlman et al. 1998). Strong polarized radio emissions, over 5\%, were detected at cm and mm wavelengths (Trippe et al. 2010; Linford et al. 2012). \astrobj{PKS 0507+17} was treated as a $\gamma$-ray candidate and included in the CGRaBS (Healey et al.2008) and CRATES (Healey et al. 2007) surveys. \astrobj{PKS 0507+17} was also included in the Owens Valley Radio Observatory (OVRO) 40 m telescope monitoring program. The program encompasses over 1500 objects, most of which are blazars, with observations for each source twice per week at a frequency of 15~GHz (Richards et al. 2011). At the infrared wavelength, \astrobj{PKS 0507+17} was included in the WISE All-sky Source Catalog and hints of intraday variability in W1 and W2 bands indicate that its infrared emission is probably dominated by the jet emission (Wright et al. 2010).

In this paper, we report a detailed analysis of the $\gamma$-ray data of \astrobj{2FGL J0509.9+1802}, especially focusing on April 2013 when extraordinary $\gamma$-ray variability has been detected by {\it Fermi}/LAT. The paper is organized as follows: In Section 2 we present the LAT data analysis. In Section 3 we report our results of $\gamma$-ray localization, $\gamma$-ray flux and spectral variability analysis. Discussions of the location of the $\gamma$-ray emission is shown in Section 4. Finally, we present the summary of our study. A cosmology model with $\Omega_{\Lambda}$=0.73, $\Omega_{M}$=0.27, and $H_{0}$=71 km $\rm s^{-1}$ $\rm Mpc^{-1}$ is adopted. Throughout this paper, we refer to a spectral index $\alpha$ as the energy index such that $F_{\nu}\propto\nu^{-\alpha}$, corresponding to a photon index $\Gamma = \alpha+1$.

\section{LAT data analysis}
The latest {\it Pass 7 Reprocessed} Data for \astrobj{2FGL J0509.9+1802} were downloaded from the LAT data server. $\gamma$-ray photon events, with time range from 4th Aug. 2008 to 6th Mar. 2014 and energy range from 0.1 to 300 GeV, are selected. The LAT data were analyzed by the updated standard {\it ScienceTools} software package version  v9r32p5 with the instrument response functions of $P7REP\_SOURCE\_V15$. For the LAT background files, we used $gll\_iem\_v05.fits$ as the galactic diffuse model and $iso\_source\_v05.txt$ for the isotropic diffuse emission template\footnote{http://fermi.gsfc.nasa.gov/ssc/data/access/lat/BackgroundModels.html}. The entire data set was filtered with {\it gtselect} and {\it gtmktime} tasks by following the standard analysis threads. Only events belonging to the class 2 were considered.

We used the unbinned likelihood algorithm (Mattox et al. 1996) implemented in the {\it gtlike} task to extract the flux and spectra. Power-law was used as the spectral template. All sources from the second-year LAT catalog (2FGL; Nolan et al. 2012) within $15^{\circ}$ of the source position were included. The flux and spectral parameters of sources within $10^{\circ}$ region of interest (ROI) were set free, while parameters of those sources that fell outside the ROI were fixed at the 2FGL values. In the fitting process, the background sources with negative TS values were removed from the source model file. Exotic parameters of background sources reaching the limits set were fixed at the 2FGL value. For the sub-daily light curve analysis, we followed the method adopted in Saito et al. (2013). For the $\gamma$-ray neighbors in the ROI, their spectral parameters were fixed in the 2FGL value, while the normalization parameters remained to be free. For the diffuse background components, their normalization parameters were frozen at the average value obtained by the analysis of the entire epoch. For each fit in the $\gamma$-ray analysis, if the test statistic (TS) value is lower than 4 or error of the flux is larger than the flux itself, the value of the flux would be replaced by 2$\sigma$ upper limits. All errors reported in the figures or quoted in the text for $\gamma$-ray flux are 1$\sigma$ statistical error.

\section{Results}
\begin{figure}
\centering
\includegraphics[scale=0.4]{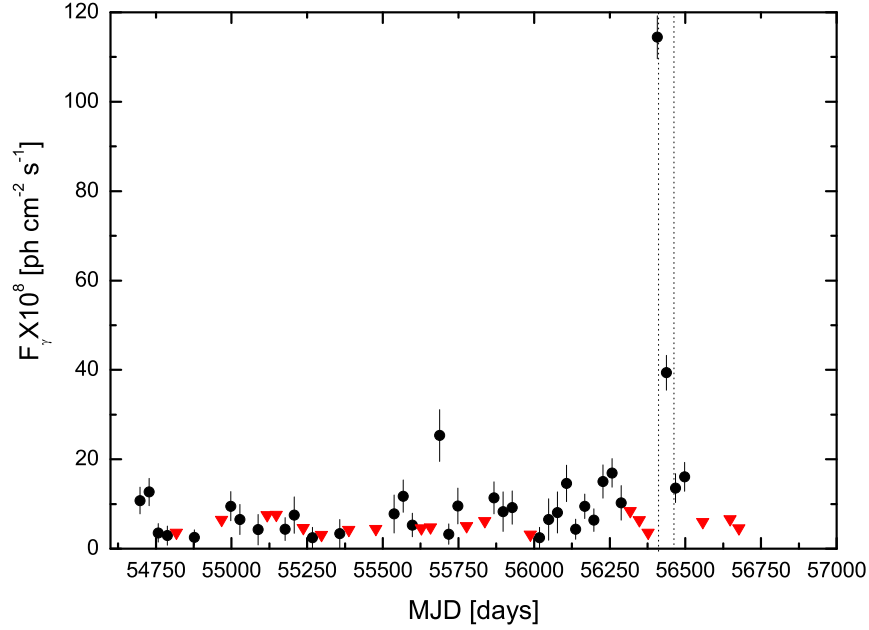}
\caption{Monthly $\gamma$-ray light curve of \astrobj{2FGL J0509.9+1802}. Red triangles represent 2$\sigma$ upper limits. The two dotted vertical lines represent two monthly time bins that the $\gamma$-ray spectra are extracted.
}
\label{Fig.1}
\end{figure}
Monthly (30 days) time bin $\gamma$-ray light curve has been extracted from the LAT data, show in Figure 1. The extraordinary $\gamma$-ray flare peaking at April 2013 is highly distinguishable. It suddenly comes out after the $\gamma$-ray low state where TS values of three time bins are lower than 4. The monthly peaking flux is $(114.4\pm4.8)\times10^{-8}$ ph $\rm cm^{-2} s^{-1}$ with TS value of 3899.4 ($\simeq62.4\sigma$). While the first two-year average flux is $(3.5\pm0.6)\times10^{-8}$ ph $\rm cm^{-2} s^{-1}$ with TS value of 133.7 ($\simeq11.6\sigma$). Since two radio counterparts correspond to the $\gamma$-ray source, it is nearly improbable that both counterparts enter to the extremely active state at the same time. During this flaring epoch, $\gamma$-ray photons are probably purely generated from a single counterpart. On the other side, large amount of received $\gamma$-ray photons is helpful to deduce the containment angle of $\gamma$-ray location. We localize the $\gamma$-ray photons of \astrobj{2FGL J0509.9+1802} during this flaring epoch from April 10th to July 9th 2013 by the task {\it gtfindsrc}. The 95\% C.L. containment angle is significantly reduced, nearly five times smaller than it of 2FGL. Only \astrobj{PKS 0507+17} remains within the 95\% C.L. containment angle, which means the $\gamma$-ray photons of \astrobj{2FGL J0509.9+1802} should correspond to \astrobj{PKS 0507+17} at this epoch.
Localizing only the Front-converting $\gamma$-ray photons whose Point-spread function is sharp confirms our result. The $\gamma$-ray emission of \astrobj{PKS 0507+17} is identified at the first time. To search the possible $\gamma$-ray emission from the other radio counterpart CRATES J0509+1806, we also localize $\gamma$-ray photons before the flare. Localization of the first 4.7 year LAT $\gamma$-ray photons suggests that CRATES J0509+1806 falls out from the 95\% C.L. containment angle. It is not surprising since CRATES J0509+1806 has been already at the edge of the 95\% C.L. containment circle in 1FGL and 2FGL. For the whole 4.7 year epoch, its contribution of $\gamma$-ray emission is probably negligible. However, for a short period, except the flare epoch, whether the $\gamma$-ray emission comes from PKS 0507+17 or CRATES J0509+1806 is actually unknown. The information of $\gamma$-ray locations and their separation angles from the radio counterparts are listed in Table~1.

The $\gamma$-ray spectra corresponding to different states have been extracted, shown in Figure 2. A
combination of a power law with an exponential cut-off gives an acceptable description to the flaring state spectrum. The photon index of the power-law component is $\Gamma$=1.82$\pm$0.02 and the break energy is $\rm E_{br}=(8.0\pm3.6)$~GeV. While the quiet state spectrum only allows us to perform a single power-law fit, with photon index of $\Gamma$=2.14$\pm$0.12. The flaring state spectrum is much harder than the quiet state one. Together with the rapid and high-amplitude $\gamma$-ray variability, these observational phenomena could be well described by the classic variability model of FSRQs that contains a injection of accelerated high energy electrons and radiative cooling process by IC scattering of the external soft photons, like the BLR emission (Sikora et al. 2001).

\begin{figure}
\centering
\includegraphics[scale=0.35]{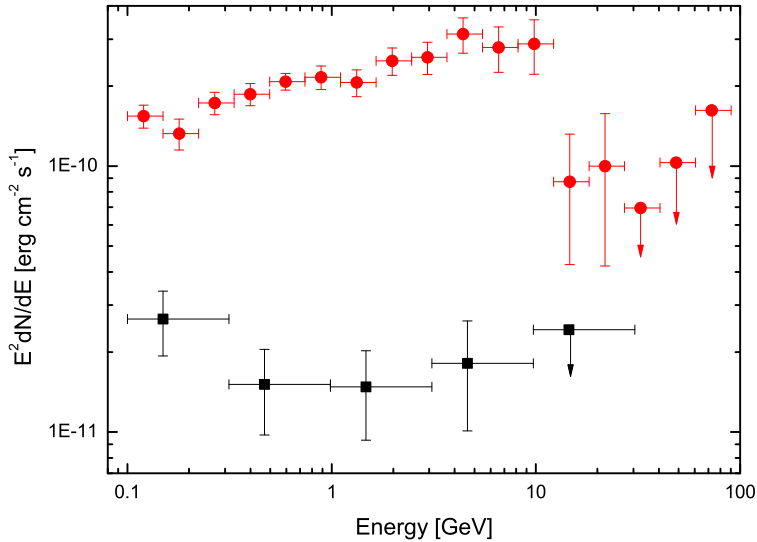}
\caption{$\gamma$-ray spectra for different states. The flaring state spectrum is shown as red filled circles, while the quiet state spectrum is shown as black squares.
}
\label{Fig.2}
\end{figure}

Since the monthly light curve exhibits intense variability at April 2013, a detailed daily light curve has been extracted to investigate the variability on short timescale, shown in the upper panel of Figure 3. The flux starts to rise from $(1.2\pm1.0)\times10^{-7}$ ph $\rm cm^{-2} s^{-1}$ at 56401.2 MJD, then remains at about $\simeq14\times10^{-7}$ ph $\rm cm^{-2} s^{-1}$ for four days, reaches the highest value at 56407.2 MJD. It takes only about 6 days to complete the rising phase, while the fading phase lasts about 27 days with several post secondary flares. The daily peak flux, $(37.9\pm4.3)\times10^{-7}$ ph $\rm cm^{-2} s^{-1}$, is over two orders of magnitude higher than the first two-year average flux, together with $\Gamma$=1.90$\pm$0.08, giving the isotropic $\gamma$-ray luminosity of $L_{\gamma,iso}=(4.2\pm1.0)\times10^{48}$ erg $\rm s^{-1}$. Adopting the relation between the jet kinetic power and the isotropic $\gamma$-ray luminosity (Nemmen et al. 2012), the jet kinetic power of \astrobj{PKS 0507+17} is therefore $P_{jet}\sim 10^{46}$ erg $\rm s^{-1}$.

\begin{figure}
\centering
\includegraphics[scale=0.4]{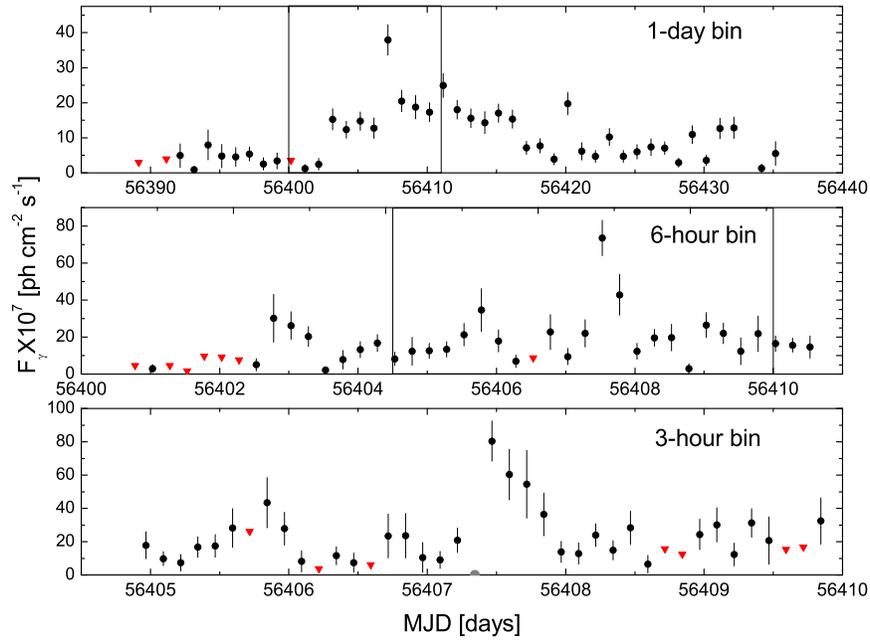}
\caption{Daily, 6-hour and 3-hour time bin $\gamma$-ray light curves of \astrobj{2FGL J0509.9+1802} around the
extraordinary $\gamma$-ray flare. The vertical lines correspond the time epoches when the more detailed light curves are extracted. Red triangles represent 2$\sigma$ upper limits. In the bottom panel, the grey symbol represents the epoch when few $\gamma$-ray photons are collected from the target area by LAT, making $\gamma$-ray flux of the target impossible to extract.
}
\label{Fig.3}
\end{figure}

Sub-daily variability has been shown in several places in the daily light curve, so we extract the 6-hour time bin light curve to derive the minimum variability timescale, shown in the middle panel of Figure 3. The classic method, $\tau_{d}=\Delta t\times ln2/ln(F_{1}/F_{2})$, is adopted to estimate the flux doubling timescale. $\gamma$-ray flux increases from $(5.0\pm3.3)\times10^{-7}$ ph $\rm cm^{-2} s^{-1}$ at 56402.53 MJD to $(30.2\pm12.9)\times10^{-7}$ ph $\rm cm^{-2} s^{-1}$ at 56402.78 MJD within 0.25 day, indicative of $\tau_{d}\simeq$2.3 hours. For another case, $\gamma$-ray flux increases from $(22.0\pm7.4)\times10^{-7}$ ph $\rm cm^{-2} s^{-1}$ at 56407.28 MJD to $(73.6\pm9.5)\times10^{-7}$ ph $\rm cm^{-2} s^{-1}$ at 56407.53 MJD within 0.25 day, indicative of $\tau_{d}\simeq$3.4 hours. Moreover, we also extract the 3-hour time bin light curve around the extraordinary $\gamma$-ray flare, shown in the bottom panel of Figure 3. Unfortunately, we do not find any evidences that support a more rapid $\gamma$-ray variation from the 3-hour bin light curve. Although several studies have been performed to search the shortest variability for FSRQs in the GeV domain, evidences of flux doubling time $\leq$ 2-3 hours have been only reported for four sources: \astrobj{PKS 1510-089}, \astrobj{3C 454.3}, \astrobj{PKS 1222+216} and \astrobj{3C 273} (Foschini et al. 2011; Vovk \& Neronov 2013). In the normal observation mode, LAT performs a complete and uniform coverage of the sky in 3 hours and 3-hour bin is the smallest bin allowed with the standard software. Since the light curves of \astrobj{PKS 0507+17} are extracted during the survey mode operation of LAT, due to the limited exposure time, doubling timescales of 2-3 hours should be treated as upper limits only. And these doubling timescales are comparable with the scan period, which means the derived timescales are not limited only by the exposure time itself.

\section{Discussion}
Since the rapid $\gamma$-ray variability is helpful to understand the location of $\gamma$-ray emission of blazars, detailed analyses for the $\gamma$-ray light curves of bright blazars observed by LAT have been performed. Flux doubling timescales on a few hours have been reported for \astrobj{PKS 1454-354} (Abdo et al. 2009), \astrobj{PKS 1502+106} (Abdo et al. 2010b), \astrobj{3C 273} (Abdo et al. 2010c), \astrobj{PKS 1222+216} and \astrobj{3C 454.3} (Tavecchio et al. 2010). Recently, GeV $\gamma$-ray doubling timescales down to 1 hour (Saito et al. 2013) and 20 minutes (Foschini et al. 2013) have been reported for \astrobj{PKS 1510-089}. And our analysis puts a new source, \astrobj{PKS 0507+17}, into this list that contains blazars with detection of $\gamma$-ray variability on a few hours. These emerging timescales on a few hours are accepted to be directly linked to scale of the emission region, $R_{em}\leq c\tau_{d}\delta(1+z)^{-1}$. By assuming a self-similar structure of relativistic jets in AGNs (Heinz \& Sunyaev 2003), distance between the emission region and the center black hole could be estimated as, $D_{em}\simeq R_{em}/\Psi$, where $\Psi$ is a scaling factor, $\Psi=1/\delta$. For \astrobj{PKS 0507+17}, assuming a typical Doppler factor value for FSRQs, $\delta$=20, and $\tau_{d}$=3 hours, the $D_{em}$ therefore is $9\times10^{16}$ cm. As the typical scale of the BLR is about 0.1 pc, $\simeq3\times10^{17}$ cm, the dissipation region of PKS 0507+17 would be located inside the BLR.

\begin{figure}
\centering
\includegraphics[scale=0.35]{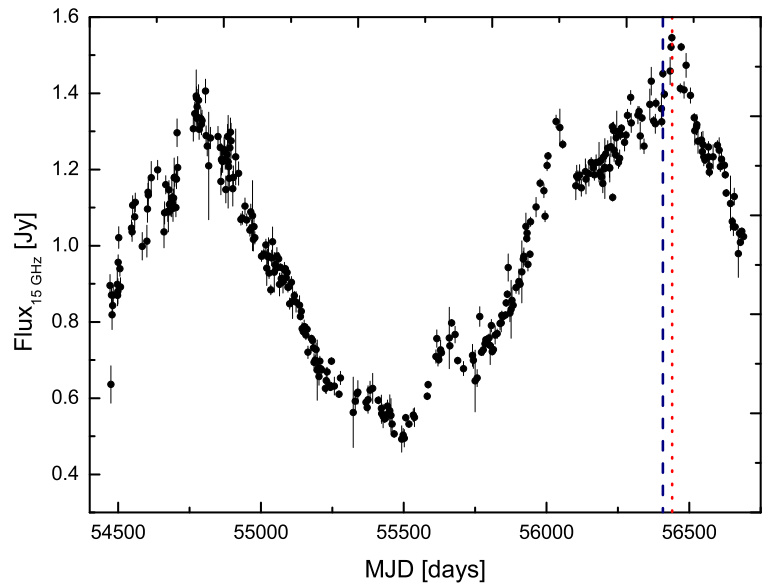}
\caption{The OVRO light curve for \astrobj{PKS 0507+17} (Richards et al. 2011). The dashed blue line corresponds to the peaking time of the  extraordinary $\gamma$-ray flare. The dotted red line correspond to the time when the radio flux is at the highest value since the operation of {\it Fermi}/LAT.
}
\label{Fig.4}
\end{figure}

Besides the rapid $\gamma$-ray variability, multi-wavelength correlation and lag analyses can also shed light on the $\gamma$-ray location. Recently, cross-correlation analysis between $\gamma$-ray and F-GAMMA radio light curves (11 cm to 0.8mm wavelength) for 54 {\it Fermi} bright blazars has been performed, suggesting that the radio emission typically lags the $\gamma$-ray emission indicative of a $\gamma$-ray location inside the radio core (Fuhrmann et al. 2014). Similar findings have been reported for two OVRO monitored blazars (Max-Moerbeck et al. 2013). Since the dissipation region of PKS 0507+17 is likely located inside the BLR, a strong radio flare that lags the extraordinary $\gamma$-ray flare is expected. \astrobj{PKS 0507+17} is included in the OVRO monitoring list, so we derive the archival 15 GHz data of \astrobj{PKS 0507+17} from The OVRO public database\footnote{http://www.astro.caltech.edu/ovroblazars/}. The 5-year 15 GHz light curve is shown in Figure 4. One of the strongest radio flare with the highest radio flux since the operation of {\it Fermi}/LAT indeed emerges about 30 days after the extraordinary $\gamma$-ray flare. However, due to the detection of only a single pair of quasi-simultaneous radio and $\gamma$-ray flares, the connection between these two spectral domains is needing a further investigation.

In summary, we have detailedly analyzed the LAT data of {2FGL J0509.9+1802} which exhibits an extraordinary $\gamma$-ray variability at April 2013. A precise localization of $\gamma$-ray emission during this epoch suggests \astrobj{PKS 0507+17} is the exclusive radio counterpart. The $\gamma$-ray emission of PKS 0507+17 is identified at the first time. The daily peak flux of the $\gamma$-ray flare is over two orders of magnitude higher than the first two-year average flux, giving an isotropic $\gamma$-ray luminosity of $\simeq4\times10^{48}$ erg $\rm s^{-1}$. Rapid $\gamma$-ray variability with doubling time of 2-3 hours has been detected by the detailed light curve analysis, indicative of a location of $\gamma$-ray emission inside the BLR. Bluer-when-brighter $\gamma$-ray spectra have been detected. These variability phenomena could be well explained by the classic FSRQ variability model that includes a fast injection of accelerated high energy electrons and the external Compton cooling process.
\section{Acknowledgments}
We are grateful to Dr. Luigi Foschini and the anonymous referee for constructive comments and suggestions. This work is financially supported by the National Natural Science Foundation of China (NSFC;11133006). This research has made use of data obtained from the High Energy Astrophysics Science Archive Research Center, provided by NASA/Goddard Space Flight Center. This research has made use of data from the OVRO 40 M Telescope Fermi Blazar Monitoring Program which is supported by NASA under awards NNX08AW31G and NNX11A043G, and by the NSF under awards AST-0808050 and AST-1109911. The authors gratefully acknowledge the computing time granted by the Yunnan Observatories, and provided on the facilities at the Yunnan Observatories Supercomputing Platform.

\clearpage

\begin{table}
\caption{$\gamma$-ray locations of \astrobj{2FGL J0509.9+1802} and angular separations from their radio counterparts. R.A./Dec. is the right ascension/declination of the $\gamma$-ray location in equatorial coordinate system (J2000). $\theta_{95}$ is the 95\% C.L. containment angle of $\gamma$-ray location. $\Delta r_{1}$ is the angular separation between $\gamma$-ray source and \astrobj{CRATES J0509+1806}, while $\Delta r_{2}$ is the separation between $\gamma$-ray source and \astrobj{PKS 0507+17} (Abdo et al. 2010a; Ackermann et al. 2011).}
\centering
\scalebox{0.9}{
\begin{tabular}{cccccc}
\hline
& R.A. & DEC. & $\theta_{95}$ & $\Delta r_{1}$ & $\Delta r_{2}$ \\
\hline
1LAC & 77.501 & 18.0157 & 0.116 & 0.115 & 0.009 \\
2LAC & 77.4933 & 18.0343 & 0.096 & 0.096 & 0.028 \\
flare epoch & 77.5084 & 18.0186 & 0.019 & 0.117 & 0.007 \\
flare epoch (front) & 77.5116 & 18.0197 & 0.021 & 0.119 & 0.008 \\
pre-flare epoch (4.7 yr) & 77.5082 & 18.0198 & 0.034 & 0.116 & 0.008 \\
\hline
\end{tabular}}
\end{table}

\end{document}